\documentclass[12pt]{iopart}

\newcommand{\pT}{$p_{\rm{t}}$~}

\newcommand{\TPC}{\rm{TPC}}

\newcommand{\VZERO}{\rm{VZERO}}

\newcommand{\cme}{\textit{Chiral Magnetic Effect}}
\usepackage{graphicx}
\begin{document}

\title[Charge dependent azimuthal correlations in ALICE]{Charge dependent azimuthal correlations in Pb--Pb collisions at $\sqrt{s_{NN}} = 2.76$~TeV}

\author{Panos Christakoglou for the ALICE Collaboration}
\address{Nikhef,
Science Park 105,
1098 XG Amsterdam,
The Netherlands}
\ead{Panos.Christakoglou@nikhef.nl}

\begin{abstract}
Separation of charges along the extreme magnetic field created in non-central relativistic 
heavy--ion collisions is predicted to be a signature of local parity violation in strong interactions.
We report on results for charge dependent two particle azimuthal correlations with respect to 
the reaction plane for Pb--Pb collisions at $\sqrt{s_{NN}} = 2.76$~TeV recorded in 2010 with 
ALICE at the LHC. The results are compared with measurements at RHIC energies and against 
currently available model predictions for LHC. Systematic studies of possible background effects 
including comparison with conventional (parity-even) correlations simulated with Monte Carlo 
event generators of heavy--ion collisions are also presented.
\end{abstract}


\vspace{-1 cm}

\section{Introduction}

The prospect of observing parity violation from the strong interaction in relativistic heavy--ion 
collisions has recently gained great attention \cite{Ref:Kharzeev}. In a highly excited dense state, 
quantum fluctuations of the gluonic field can produce configurations with different local parities. 
The interaction of these local bubbles of chiral symmetry restoration and non-vanishing topological 
charges, with the strong and short--lived magnetic field produced in non--central heavy--ion collisions 
\cite{Ref:BField1} gives rise to the \cme. The effect is argued to be reflected in the asymmetry in the 
emission of particles relative to the reaction plane: opposite charged particles will be emitted 
preferentially in different directions across the reaction plane.

In this article, we report the results for the two-- and three--particle charge dependent azimuthal 
correlations with respect to the reaction plane for Pb--Pb collisions at $\sqrt{s_{NN}} = 2.76$~TeV 
recorded in 2010 with ALICE  \cite{Ref:ALICE} at the LHC. The tools used for this study are the 
2-- and 3--particle azimuthal correlators represented by $\langle \cos(\phi_{\alpha} - \phi_{\beta}) \rangle$ 
and $\langle \cos(\phi_{\alpha} + \phi_{\beta} - 2\Psi_{RP}) \rangle$ 
\cite{Ref:Sergey3particleCorrelator}, respectively.

\begin{figure}[h!]
\includegraphics[height=7.1cm,width=16cm]{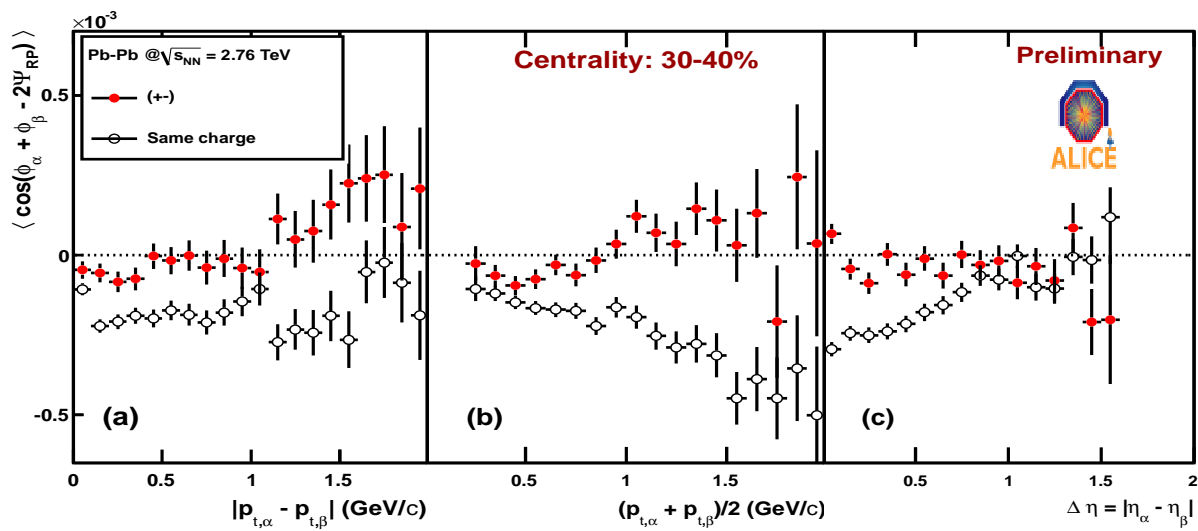}
\caption{[Color online] The dependence of the 3--particle correlator on the: (a)  \pT difference of the 
pair; (b) average \pT of the pair; (c) $\Delta \eta$ of the pair. The legend indicates the connection 
between the charge combinations and the different marker styles used.}
\label{fig:ALICEResults} 
\end{figure}
\vspace{-0.3 cm}

\section{Results}

Data from around 50~M Pb--Pb events at $\sqrt{s_{NN}} = 2.76$~TeV were analyzed. Only events with 
a proper online trigger were considered in the analysis. An offline event selection was also applied to 
reduce the contamination from background events. Events having a reconstructed vertex were used in 
this study. The centrality of the collision was estimated by using the distribution of the signal from the 
\VZERO~scintillator detectors, placed around the beam pipe on either side of the interaction region and 
covering the pseudorapidity ranges$2.8 < \eta < 5.1$ and  $-3.7 < \eta < -1.7$ \cite{Ref:ALICE}, and 
fitting it with a Glauber model. The tracks were reconstructed by the main tracking device of ALICE, the 
\TPC~\cite{Ref:ALICE}, that provides a uniform acceptance with minimal corrections. Adequate selection 
criteria were applied to minimize the contribution from background tracks. Finally, the phase space 
analyzed was given by $|\eta| < 0.8$ and $0.2 < p_{\rm{t}} < 5.0$~GeV/c.

Figure \ref{fig:ALICEResults} shows the dependence of the 3--particle correlator on the \pT difference, 
the mean \pT and the pseudo--rapidity difference of the pair, in the left, middle and right plots, respectively. 
The black circles correspond to pairs having the same charge,  while the red squares representing the 
correlations between opposite charges. In all plots, the 30--40$\%$ centrality percentile is shown. The 
opposite charged pairs don't show any dependence on either of the three variables used and their values 
are consistent with 0. On the other hand, the same charged pairs demonstrate a different pattern: the data 
points don't indicate a significant contribution from short range correlations (left plot), the 
magnitude of the correlations seems to increase with increasing mean \pT of the pair (middle plot), 
whereas the $\Delta \eta$ dependence shows a width of one unit in $\Delta \eta$.

\begin{figure}[h!]
\includegraphics[height=7.1cm,width=8cm]{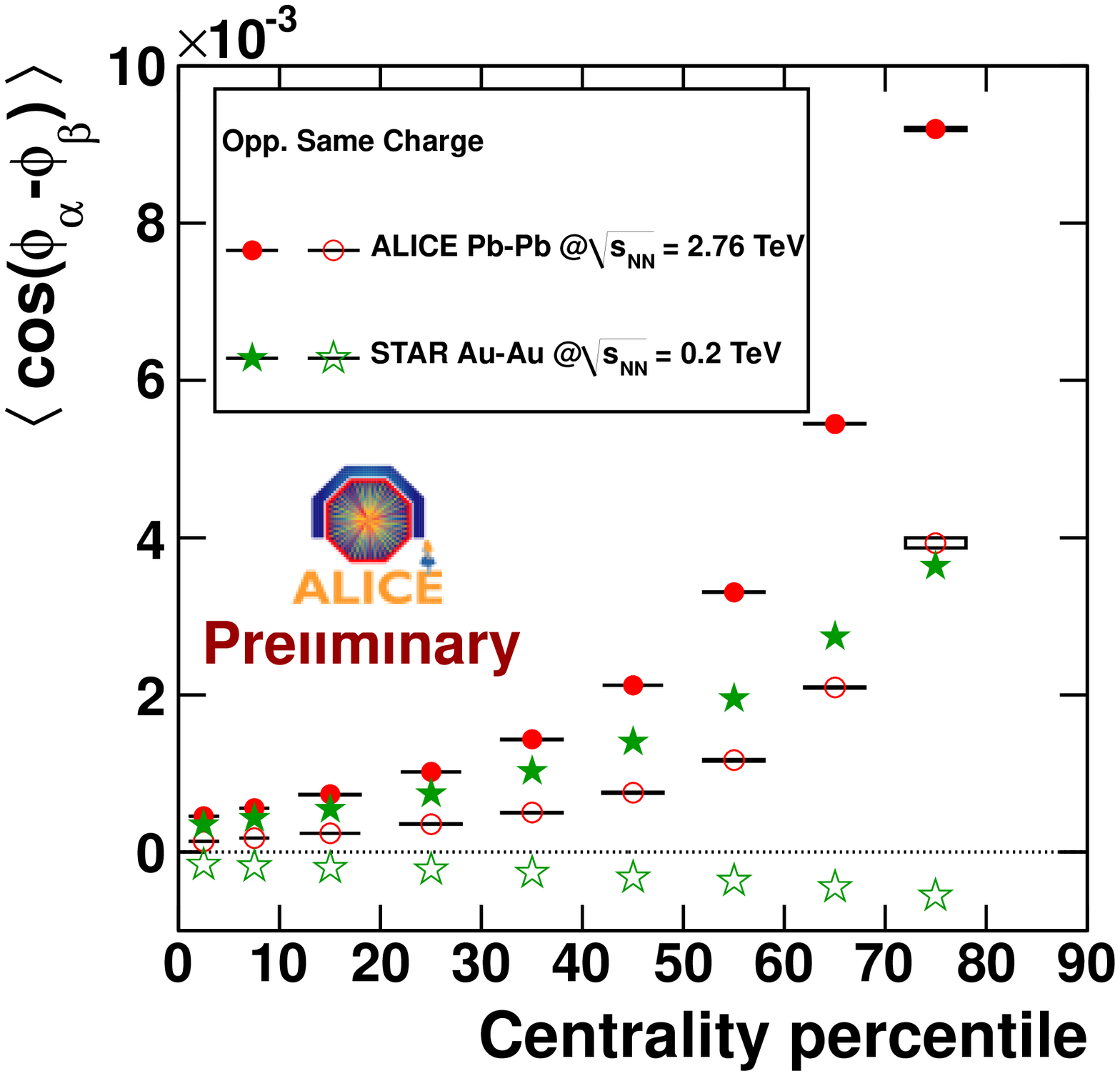}
\includegraphics[height=7.1cm,width=8cm]{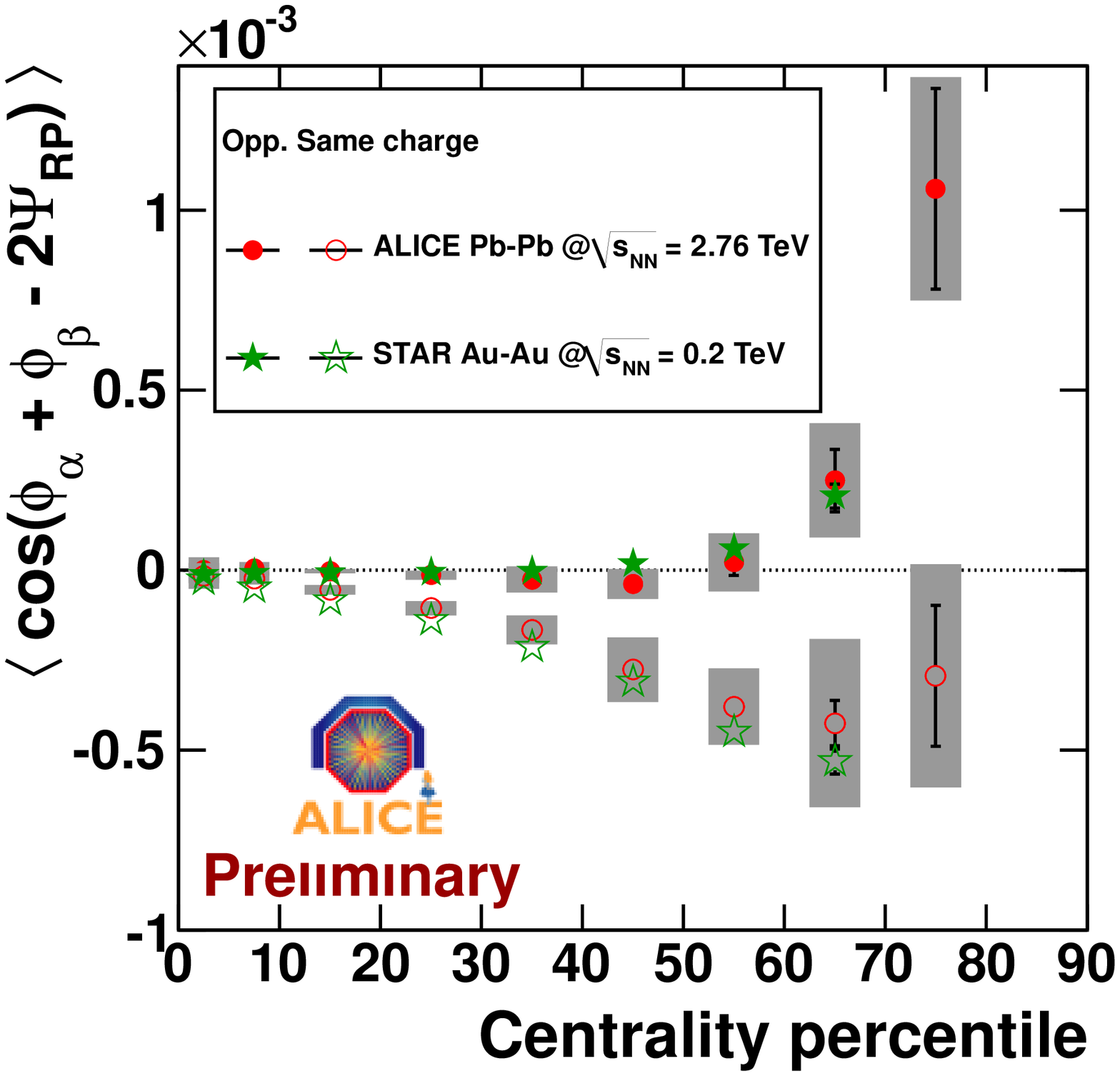}
\caption{[Color online] The comparison of the integrated 2-- (left plot) and 3--particle correlator (right plot) 
between ALICE and STAR. The legend indicates the connection between the charge combinations and 
the different color and marker styles used.}
\label{fig:ALICEvsSTAR} 
\end{figure}

Figure \ref{fig:ALICEvsSTAR} shows the centrality dependence of the integrated 2-- (left plot) and 
3--particle (right plot) correlators measured by ALICE in Pb--Pb collisions at $\sqrt{s_{NN}} = 2.76$~TeV 
(red circles) compared to the values reported by STAR in Au--Au collisions at $\sqrt{s_{NN}} = 0.2$~TeV 
(green stars) \cite{Ref:STARParity}. In both plots, the error bars correspond to the statistical uncertainties 
whereas the systematic ones are represented by the shaded areas. The full markers correspond to the 
correlations between oppositely charged pairs, with the open ones showing the same charged pairs. 
The 2--particle correlation analysis indicates that there is a change of sign in the correlations of same 
charged pairs between the LHC and RHIC energies, indicating in turn a change of the correlation pattern 
in-- and out--of--plane in the two experiments. This manifests itself by the fact that the results for the 
correlations of same charged pairs at RHIC energies have the same magnitude when measured with 
the 2-- and 3--particle correlation technique: at RHIC energies particles exhibit a stronger magnitude of 
their correlations in the direction along than across the reaction plane. The results obtained from the 
3--particle correlation analysis  show a clear charge separation when moving to more peripheral 
collisions. There is a remarkable agreement  between the measurements at two different energies. 
Combining both correlation techniques for ALICE, we conclude that the relevant correlations have 
stronger magnitude out--of than in--plane.

\begin{figure}[h!]
\begin{center}
\includegraphics[height=7.1cm,width=14cm]{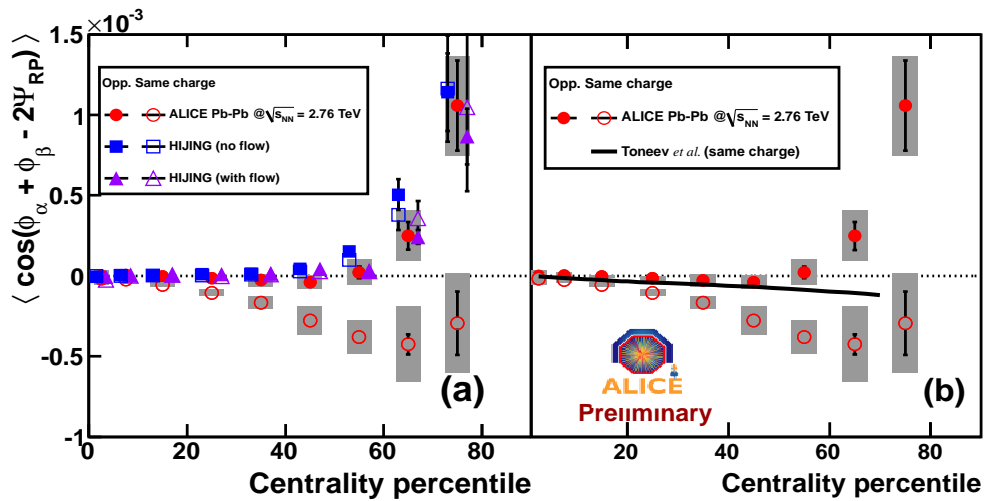}
\caption{[Color online] (a) Comparison of the centrality dependence of the integrated 
3--particle correlators $\langle \cos(\phi_{\alpha} + \phi_{\beta} - 2\phi_{c}) \rangle$ 
between data and HIJING \cite{Ref:HIJING}. (b) Comparison of the centrality dependence of the 
measured 3--particle correlator with theory predictions.}
\label{fig:ALICEComparison} 
\end{center}
\end{figure}

Figure \ref{fig:ALICEComparison}--left presents the comparison of the centrality dependence of the 
integrated correlator $\langle \cos(\phi_{\alpha} + \phi_{\beta} - 2\Psi_{RP}) \rangle$ 
\cite{Ref:Sergey3particleCorrelator}, between the experimental points and HIJING \cite{Ref:HIJING}. 
The blue squares represent the analysis of HIJING events without any after--burner. In addition, an 
after--burner was also used to modify the relevant distributions of generated particles with the realistic 
values of the differential elliptic flow for each centrality bin \cite{Ref:AliceFlow}. The corresponding points 
are represented by the purple triangles. The correlations between opposite charged pairs are described 
well by the HIJING points, the latter exhibiting no major differences related to the usage of the after--burner. 

Figure \ref{fig:ALICEComparison}--right shows the comparison of the experimental data (red points) with 
predictions from theory \cite{Ref:Toneev} for LHC energies. The blue line shows the prediction for the 
centrality dependence of the correlations between same charged pairs, assuming the existence of the 
\cme, with a certain value for the starting time of the evolution of the magnetic field. This particular 
model clearly underestimates the experimental result. The experimental observation of no apparent 
change in the magnitude of the effect between RHIC and LHC contradicts the quantitative predictions. 
On the other hand, a hint that the signal might be similar at different energies was already given in 
\cite{Ref:KharzeevEnergy}.

\vspace{-0.3 cm}
\section{Summary}

We presented the first measurement of the 2-- and 3--particle correlators at LHC energies 
with the ALICE experiment. The results indicate that there is a change in the correlation pattern 
(i.e. larger out--of-- than in--plane correlations) between LHC and RHIC energies. The results of the 
integrated 3--particle correlator measured by ALICE show a remarkable agreement with the published 
data from STAR.

\end{document}